\documentclass[%
 aip,
 amsmath,amssymb,
 reprint,%
] {revtex4-2}

\usepackage{graphicx}
\usepackage{dcolumn}
\usepackage{bm}
\usepackage[T1]{fontenc}

\begin{document}

\preprint{Appl. Phys. Lett.}

\title[Electronic topological transition in  $\beta$-Ag$_2$Se]{Temperature induced first order electronic topological transition in $\beta$-Ag$_2$Se}

\author{L. S. Sharath Chandra}
\email{lsschandra@rrcat.gov.in}
 \affiliation{Free Electron Laser Utilization Laboratory, Raja Ramanna Centre for Advanced Technology, Indore - 452 013, India.}
\author{SK. Ramjan}
\affiliation{Free Electron Laser Utilization Laboratory, Raja Ramanna Centre for Advanced Technology, Indore - 452 013, India.}%
\affiliation{Homi Bhabha National Institute, Training School Complex, Anushakti Nagar, Mumbai 400 094, India.}%
\author{Soma Banik}
\affiliation{Electron Spectroscopy \& Materials Laboratory, Synchrotrons Utilization Section, Raja Ramanna Centre for Advanced Technology, Indore - 452 013, India.}
\affiliation{Homi Bhabha National Institute, Training School Complex, Anushakti Nagar, Mumbai 400 094, India.}%
\author{Archna Sagdeo}
\affiliation{Hard X-ray Applications Laboratory, Synchrotrons Utilization Section, Raja Ramanna Centre for Advanced Technology, Indore - 452 013, India.}%
\affiliation{Homi Bhabha National Institute, Training School Complex, Anushakti Nagar, Mumbai 400 094, India.}%
\author{M. K. Chattopadhyay}
\affiliation{Free Electron Laser Utilization Laboratory, Raja Ramanna Centre for Advanced Technology, Indore - 452 013, India.}%
\affiliation{Homi Bhabha National Institute, Training School Complex, Anushakti Nagar, Mumbai 400 094, India.}%

\date{\today}

\begin{abstract}

$\beta$-Ag$_2$Se is a promising material for room temperature thermoelectric applications and magneto-resistive sensors. However, no attention was paid earlier to the hysteresis in the temperature dependence of resistivity ($\rho$($T$)). Here, we show that a broad hysteresis above 35~K is observed not only in $\rho$($T$), but also in other electronic properties such as Hall coefficient ($R_H$($T$)), Seebeck coefficient, thermal conductivity and ultraviolet photoelectron spectra (UPS). We also show that the hysteresis is not associated with a structural transition. The $\rho$($T$) and $R_H$($T$) show that  $\beta$-Ag$_2$Se is semiconducting above 300~K, but metallicity is retained below 300~K. While electronic states are absent in the energy range from the Fermi level ($E_F$) to 0.4~eV below the $E_F$ at 300~K, a distinct Fermi edge is observed in the UPS at 15~K suggesting that the $\beta$-Ag$_2$Se undergoes an electronic topological transition from a high temperature semiconducting state to a low temperature metallic state. Our study reveals that a constant and moderately high thermoelectric figure of merit ($ZT$) in the range 300-395~K is observed due to the broad semiconductor to metal transition in $\beta$-Ag$_2$Se.  
\end{abstract}

\maketitle

The chalcogenides are considered to be promising materials for thermoelectric applications \cite{Jaz20, hos20, wol19, Mao18, he17, ish13}. Ag$_2$Se is especially attractive due to it's high figure of merit ($ZT$) near room temperature \cite{fer00, day13} and the abundance of it's constituent elements on earth's crust \cite{day13}. The $\beta$-Ag$_2$Se is also found promising for high field magneto-resistive sensor applications \cite{hus02}. Ag$_2$Se undergoes a structural transition at $T_{\beta-\alpha}$ = 407~K from a low temperature semiconducting (space group: C2/m) orthorhombic $\beta$-phase to a high temperature superionic face centred cubic $\alpha$-phase (space group: Fm$\bar{3}$m) \cite{dry13}. The $ZT$ of $\beta$-Ag$_2$Se near 300~K varies from 0.2 to 1.0 depending on the charge carrier concentration ($n$) \cite{day13}. The optimized $n$ = 1.6 $\times$ 10$^{18}$~cm$^{-3}$ for the maximum $ZT$ in  $\beta$-Ag$_2$Se is small in comparison with other chalcogenides \cite{day13}. This creates a challenge for the use of $\beta$-Ag$_2$Se based materials in thermoelectric applications as the ionic movement of Ag above $T_{\beta-\alpha}$ results in the segregation of Ag clusters along the grain boundaries \cite{kum96}, thereby changing the dynamics of the system. The temperature dependence of resistivity $\rho$($T$) of $\beta$-Ag$_2$Se is semiconducting in the range  320~K$ < T < T_{\beta-\alpha}$ \cite{fer00,hus02,xu97,zha17,lee07,wan11, ali08,jaf10,day13,dry13,sau19,zho14,mi14,kum96}. On the other hand, $\rho$($T$) in the range 2-320~K shows a hump at $T_{max}$ with metallic behaviour (or positive temperature coefficient of resistivity) below $T_{max}$  \cite{fer00,hus02,xu97,zha17,lee07,wan11} . The position of $T_{max}$ reported by various authors differ significantly viz., 60~K \cite{xu97, wan11}, 80~K\cite{hus02,wan11}, 100~K\cite{zha17}, and 200~K \cite{fer00}. Such a behaviour in the temperature dependence of resistivity of degenerate semiconductors can be explained by considering a semiconducting band and a metallic impurity band \cite{sha08}. However, Santhosh Kumar and Pradeep \cite{san02} reported that the  $\rho$($T$) of $\beta$-Ag$_2$Se thin film during the warming cycle does not match with the cooling cycle in the range  300~K$ < T < T_{\beta-\alpha}$ leading to a large hysteresis. Indication of this hysteresis in the $\rho$($T$) of $\beta$-Ag$_2$Se thin films was also observed by Damodara Das and Karunakaran \cite{dam89}. Thermal hysteresis in these results indicate that there is a first-order phase transition \cite{chi00} from  a semiconducting state above $T_{max}$ to a metallic/semi metallic state  below $T_{max}$. However, no studies are found in the literature on the origin of such a hysteresis. Therefore, understanding the electronic behaviour of Ag$_2$Se is important for utilizing this system for technological applications at room temperature and below. 

Here we show that there is a broad hysteresis above 35~K in the electronic properties such as $\rho$($T$), charge carrier density $n$($T$), Seebeck coefficient $S$($T$), thermal conductivity $\kappa$($T$) and ultraviolet photoelectron spectra (UPS). However, there is no structural transition in the range 80 - 395~K. The UPS at T~=~15~K shows a clear Fermi edge, in comparison with the absence of electronic states down to 0.4~eV below the Fermi level at 300~K. Our results suggest that the $\beta$-Ag$_2$Se undergoes a temperature-dependent electronic topological transition.     

\begin{figure}
\includegraphics[width = 90mm]{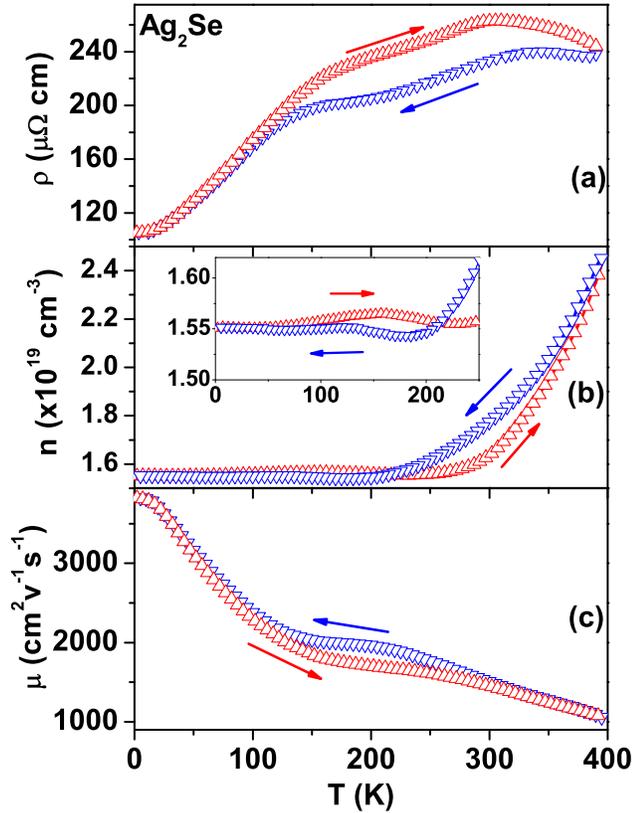}
\vskip -1.5 cm
\caption{\label{fig:epsart} (color online) Temperature dependence of (a) resistivity (b) carrier density (c) mobility of Ag$_2$Se  in the range 2-395~K. A thermal hysteresis is observed in the range 30-395~K. }
\vskip -0.5 cm
\end{figure}    

\begin{figure}
\includegraphics[width = 90mm]{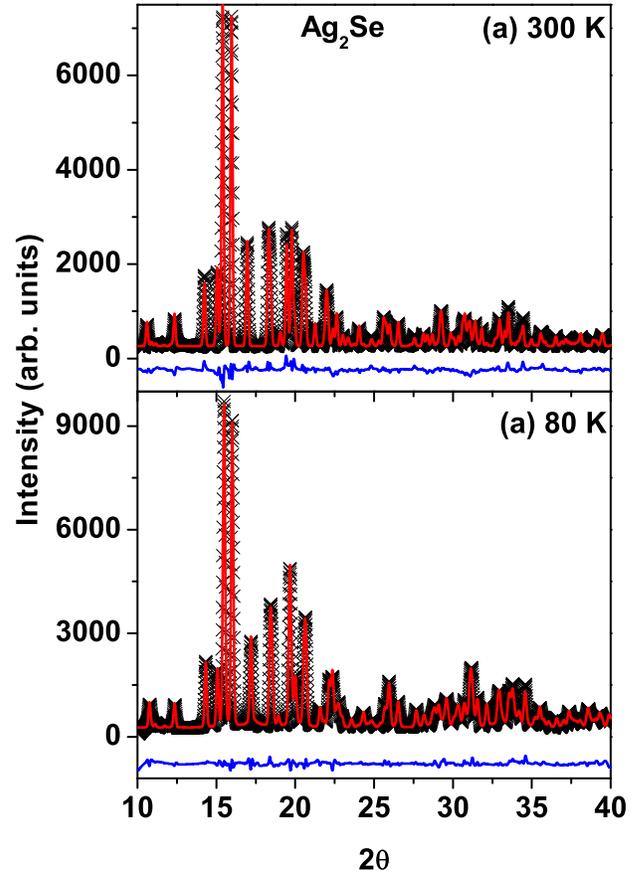}
 \vskip -0.5 cm  
\caption{\label{fig:epsart} (color online) X ray diffraction patterns of Ag$_2$Se at (a) 300~K and (b) 80~K. No change in the crystal structure is observed. The thermal expansion  along the $a$ direction is found to be negative.}
\vskip -0.5 cm
\end{figure}

Figure 1 show the (a) $\rho$($T$), (b) $n$($T$) and (c) mobility $\mu$($T$) measured using a 9~T Physical Property Measurement System (PPMS, Quantum Design, USA) in the range 2-395~K. The Ag$_{2}$Se sample was prepared by melting elemental silver (Sigma Aldrich, 99.9\%) and selenium (Otto Chemi, 99.999\%) in stoichiometric proportions in an induction furnace in He (99.9999\%) atmosphere. The $\rho$($T$) increases when the temperature decreases from 395~K to 320~K showing a semiconducting behaviour. The curvature of $\rho$($T$) changes with further decrease in temperature and shows a broad double-hump feature in the range 300~K-125~K. Below 125~K, the $\rho$($T$) decreases with decrease in $T$ like that of a metallic or semi-metallic system. The $\rho$($T$) while warming matches with that during cooling in the range 2-35~K. Above 35~K, the $\rho$($T$) exhibits a thermal hysteresis (between warming and cooling) that seems to extend above 395~K (maximum $T$ for the present measurements) indicating that it probably ends near $T_{\beta-\alpha}$. We have observed such a behaviour in many of the Ag$_2$Se and doped Ag$_2$Se alloys as well (not shown here). The $\rho$($T$) of this sample has the lowest value in comparison with that reported for Ag$_2$Se in the literature \cite{fer00,hus02,xu97,zha17,lee07,wan11, ali08,jaf10,day13,dry13,sau19,zho14,mi14,kum96}. The energy gap estimated from the $\rho$($T$) in the range 320-395~K of the heating cycle is approximately 0.02~eV, which is similar to that reported in the literature \cite{lee07, dal67, dal67a, sim63, dam89, bae62}. The $n$($T$) (Fig. 1(b)) estimated from Hall resistivity also shows thermal hysteresis. However, the hysteresis in $n$($T$) is less prominent than that of $\rho$($T$). When the temperature is decreased from 395~K, the $n$($T$) decreases and becomes nearly constant below 200~K. In the heating cycle $n$($T$) remains constant up to 270~K and then increases with increasing temperature. Though, $n$($T$) is nearly constant below 250~K , the expanded view shown in the inset to Fig.1(b)  depicts the correspondence between $\rho$($T$) and $n$($T$). The $n$($T$) shows a crossover below 250~K which is absent in $\rho$($T$). This seems to be related to the fact that the changes in $n$($T$) is dependant on the changes in the density of states, whereas, resistivity depends not only on $n$($T$), but also on the scattering of carriers from the defects. The energy gap estimated from the $n$($T$) in the range 360-395~K is approximately 0.06~eV, which is similar to that reported in literature \cite{lee07, dal67, dal67a, sim63, dam89, bae62}. The slightly lower energy gap estimated from $\rho$($T$) in comparison to that from $n$($T$) can be attributed to the temperature dependence of the relaxation of free carriers. The $\mu$($T$) is estimated from the $\rho$($T$) and $n$($T$) using the single band picture as $\mu$($T$) = $\sigma$($T$)/$n$($T$)$e$, where $e$ is the electronic charge. The $\mu$($T$) (Fig.1(c)) is $\sim$1500~cm$^{2}$V$^{-1}$s$^{-1}$ at room temperatures and reaches to about 3800~cm$^{2}$V$^{-1}$s$^{-1}$ at 2~K. The slope of the $\mu$($T$) changes at temperatures where the $\rho$($T$) shows humps. The difference in the temperature dependences of $\rho$($T$), and $n$($T$) is responsible for smaller hysteresis and the absence of crossover between heating and cooling cycles in the $\mu$($T$).  The mean free path of electrons\cite{ash00} at 2~K estimated using the experimental $n$ is about 2000~\AA~which is quite large in comparison to unit cell dimensions \cite{zha17}. Therefore, the origin of metallicity is not related to static and/or dynamic disorder.

\begin{table}[h]
	
	\begin{center}
			
			\caption{Results of Rietveld refinement of the XRD patterns of Ag$_2$Se at 300~K and 80~K}
			\label{tab:table1}
			\begin{tabular}{c|c|c} 
				\hline
				\hline
				\multicolumn{2}{c}{Sample}& Ag$_2$Se\\
				\multicolumn{2}{c}{Wavelength}& 0.7172~\AA\\
				\multicolumn{2}{c}{Crystal structure}& Orthorhombic\\
				\multicolumn{2}{c}{Space Group}& P 2$_1$ 2$_1$ 2$_1$\\
				\hline
				Parameters& {\bf300~K} & {\bf80~K}\\
				\hline
				\multicolumn{3}{c}{Lattice parameters (\AA~)}\\
				\hline
				$a$ &4.3357(1) & 4.3617(1)\\
				$b$ &7.0717(2)&7.0329(1)  \\
				$c$ &7.7786(2) &7.6648(2))  \\
				\hline
				\multicolumn{3}{c}{Wyckoff positions}\\
				\hline
				\multicolumn{3}{c}{Ag1 at }\\
				\hline
				$x$ &0.1489(5) & 0.1567(3)\\
				$y$ &0.3849(3) &0.3834(2) \\
				$z$ &0.9517(3) &0.9520(2)  \\
				\hline
				\multicolumn{3}{c}{Ag2 at }\\
				\hline
				$x$ &0.4723(4) & 0.4792(3)\\
				$y$ &0.2219(3) &0.2273(2)  \\
				$z$ &0.6391(3) &0.6319(2)  \\
				\hline
				\multicolumn{3}{c}{Se at }\\
				\hline
				$x$ &0.1119(6) & 0.1155(4)\\
				$y$ &0.0058(4) &0.0047(3)  \\
				$z$ &0.8507(4) &0.8437(2)  \\
				\hline
				\multicolumn{3}{c}{R factors and goodness of the fit }\\
				\hline
				R$_{exp}$ &9.04\% & 6.59\%\\
				R$_{wp}$ &9.4\% &6.66\%  \\
				$\chi^2$ &1.08 &1.02  \\
				\hline
				\hline
				
			\end{tabular}
			
			
			
	\end{center}
\end{table}

A change of slope in the temperature dependence of different observables (viz., Fig. 1) associated with a thermal hysteresis indicates the presence of a first-order phase transition (FOPT) (see Ref.[\cite{cha08, roy04}] and the references therein). This hysteresis originates from the metastability (superheating \& supercooling) \cite{chi00} across a FOPT, and is a distinct feature of such a phase transition \cite{chi00, cha08, roy04}. The FOPT in Ag$_2$Se seems to have been smeared out due to the influence of disorder \cite{cha10, imr79}. FOPTs observed in several materials can be ascribed to changes in crystal, magnetic and/or electronic structure \cite{roy04}. The Ag$_2$Se at low temperatures can form in different crystal structures \cite{pri20, pri20a}. The stable phase for the bulk samples is the orthorhombic structure. Other structures such as monoclinic or tetragonal are stabilized only in thin films or nanocrystals of Ag$_2$Se [Ref. \cite{pri20, pri20a} and the references therein]. There can be metastable nanocrystalline monoclinic phase in the bulk samples of Ag$_2$Se which has very low $n$ \cite{pri20, pri20a}. The $n$ of the present sample is one order magnitude higher than those containing metastable inclusions \cite{pri20, pri20a}.  

The x-ray diffraction (XRD) patterns (Fig.2) of Ag$_2$Se  at (a) 300~K and (b) 80~K measured at the angle dispersive XRD beamline BL-12 of Indus-2 Synchrotron Source\cite{sin00} are similar. The red solid line is the Rietveld refinement of the XRD patterns. The refinement parameters are listed in table I. At 300~K, the observed $a$ is smaller, and $b$ and $c$ are larger than the corresponding lattice parameters reported in the literature \cite{zha17}. This may be due to the disorder present in the sample. One can see that there is a negative thermal expansion along the $a$ direction. We could not find any indication of structural transformation or any reflections for secondary phases in the Ag$_2$Se sample from 80~K to 405~K. The changes in the atomic positions are also negligible. Therefore, no secondary phase grows to achieve percolation threshold to exhibit a change from semiconducting to metallic character. On the other hand, the metastable nanocrystalline phases are semiconducting with higher band gap \cite{pri20, pri20a}. Since, our sample is metallic/semimetallic at low temperatures, the possibility of structural origin of the observed  hysteresis in the transport properties can be negated. The magnetic moment of Ag$_2$Se is hardly measurable and no magnetic transition is observed in the temperature dependence of magnetization in the range 2-395~K (not shown here). Thus, the thermal hysteresis in $\rho$($T$) and \ $n$($T$) (Fig.1) has an electronic origin. 

\begin{figure}
\includegraphics[width = 90mm]{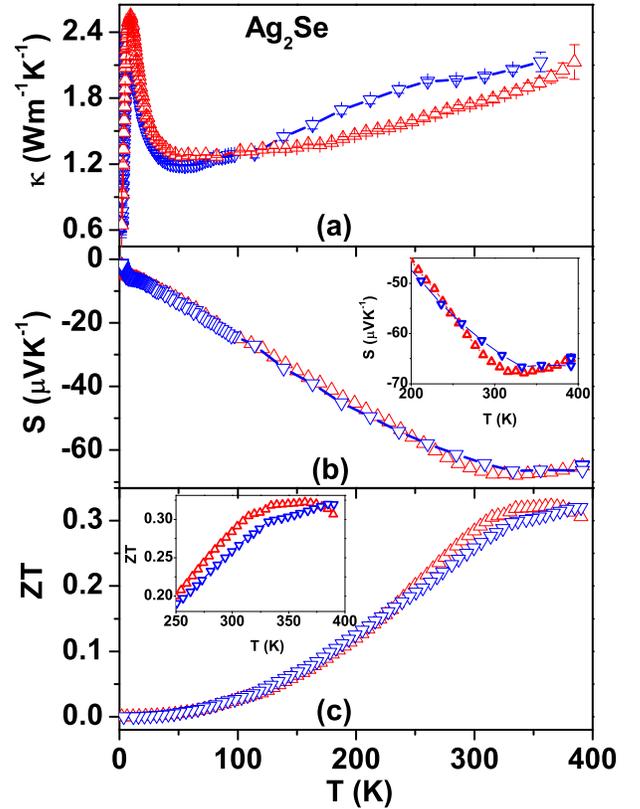}
\vskip -1.5 cm
\caption{\label{fig:epsart} (color online) Temperature dependence of (a) thermal conductivity (b) thermoelectric power (c) dimensionless figure of merit of $\beta$-Ag$_2$Se in the range 2-395~K. Both thermal conductivity and thermoelectric power exhibit thermal hysteresis. Inset to (b) shows the expanded view of the thermoelectric power near room temperature. The $ZT$ $\geq$ 0.2 (inset to (c)) for T $\geq$ 250~K.}
\vskip -0.5 cm
\end{figure}

Figure 3 shows the (a) $\kappa$($T$), (b) $S$($T$) and (c) the dimensionless figure of merit $ZT$ = $S^2T/\rho\kappa$ in the temperature range 2-395~K measured using the PPMS. The $\kappa$($T$) (Fig.3(a)) decreases with decreasing temperature in the range 50-395~K and then shows a peak at about 10~K. A quasilinear dependence of $\kappa$($T$) above 100~K indicates that majority of the heat is carried by electrons and the heat flow is limited by the scattering of electrons from the static defects \cite{trit00}, which is similar to that observed in literature \cite{mi14, wan11}. The hysteresis in the $\kappa$($T$) above 100~K is also an indication that the origin of hysteresis is electronic in nature. The Seebeck coefficient $S$($T$) (Fig. 3(b)) is nearly constant at -65~$\mu$V/K above 300~K. The absolute value of $S$($T$) decreases with decreasing temperature below 300~K. Similar to $n$($T$), the thermal hysteresis in the $S$($T$) is smaller (inset to Fig.3(b)) than that of thermal conductivity. The absolute value of $S$($T$) of the present sample is half of that reported in the literature \cite{fer00, lee07, mi14}. This is because of the high carrier concentration present in the sample. Figure 3(c) shows the $ZT$ as a function of temperature for heating and cooling cycle. The value of $ZT >$ 0.2 for $T >$ 200~K. One can also see that $ZT$ is constant at approximately 0.32 in the range 300-395~K. Thus, the broadened phase transitions may be useful in generating large constant $ZT$ over a wide temperature range.

\begin{figure}
\includegraphics[width = 90mm]{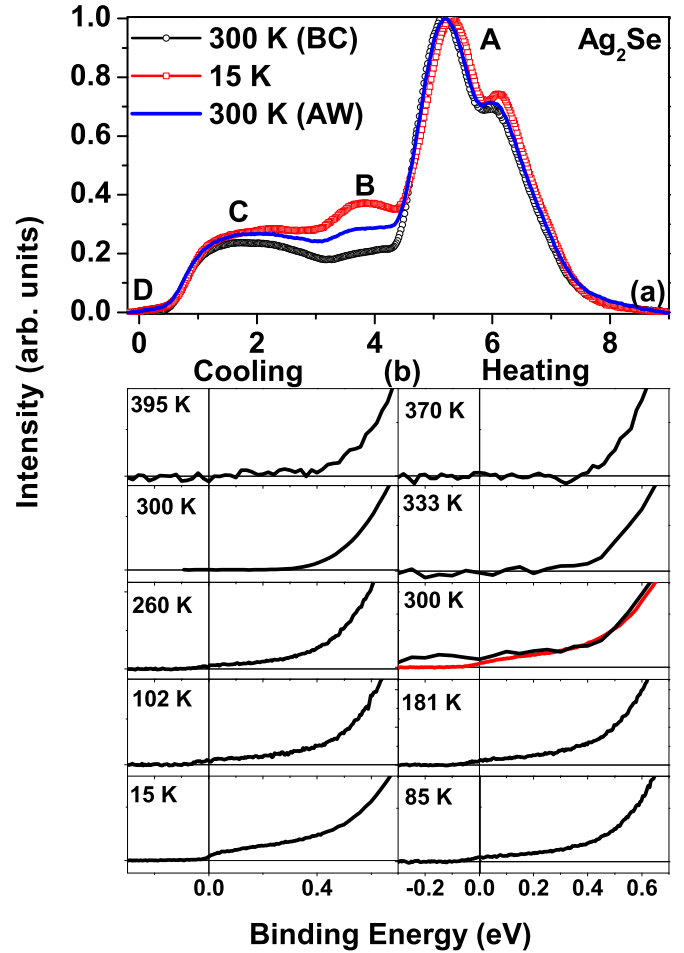}
\caption{\label{fig:epsart} (color online) (a) Ultraviolet photoemission spectroscopy of Ag$_2$Se sample at 300~K before cooling to 15~K (300~K (BC)), at 15~K and at 300~K after warming from 15~K (300~K (AW)). The valence band at 300~K is not the same before and after cooling to 15~K. (b) Expanded view of the valence band spectra near the Fermi level at different temperatures. After warming back, the electronic structure is recovered above 333~K.}
\vskip -0.5 cm
\end{figure}

An electronic transition without a structural transition when the pressure, composition and/or magnetic field are varied is called the electronic topological transition (ETT) \cite{lif60,vol17}. Recently, we have shown that drastic changes in the superconducting and normal state properties of the Mo$_{1-x}$Re$_x$ alloys are observed when the composition exceeds the critical concentrations corresponding to two ETTs at $x_{C1}$ = 0.05 and $x_{C2}$ = 0.11 \cite{shy15,shy16,sha20}. Temperature driven ETT has been observed in many systems such as Fe$_{1.08}$Te \cite{myd17},  ZrTe$_5$ \cite{chi17, xu18, zha17a}, ZrSiSe \cite{che20}, MoTe$_2$ \cite{ber18}, WTe$_2$ \cite{wu15}, and Mn$_3$Ge \cite{wan21} etc. The broad hysteresis in resistivity as a function of temperature  across the ETT  in Fe$_{1.08}$Te under pressure is reported to be due to the first-order nature of the transition \cite{myd17}. Mydeen et al., have shown that the change in $c/a$ ratio from 1.59 to 1.63 as a function of temperature in Fe$_{1.08}$Te under 2.9~GPa results in the disappearance of four Fermi sheets. This is due to the anisotropic deformation induced by the coupling of electronic states of Fermi pockets to the thermally excited acoustic phonons \cite{myd17}. Generally, the $\mu$($T$)  of $\beta$-Ag$_2$Se shows $T^{-0.5}$ dependence \cite{mi14}. In our sample, such a temperature dependence is observed in a narrow temperature range 60-150~K. The $T^{-0.5}$ dependence of $\mu$($T$) is a signature of the electron-acoustic phonon (e-ph) interaction. Temperature dependence of resistivity of non magnetic metals and degenerate semiconductors arises from e-ph interaction. Therefore, the origin of the FOPT can be ascribed to the e-ph interaction. The effect of e-ph interaction on Hall resistivity and $S$($T$) is insignificant in metals and degenerate semiconductors. Therefore, the hysteresis in $n$($T$) and $S$($T$) are quite smaller than those of $\rho$($T$) and $\kappa$($T$). The temperature-dependent XRD of Ag$_2$Se shows that the lattice parameter $b$ is nearly independent of temperature. However, $a$ increases and $c$ decreases when the temperature is decreased below $T_{\beta-\alpha}$ indicating a temperature induced anisotropic distortion in the unit cell of Ag$_2$Se. Therefore, the origin of the transition from high temperature semiconducting phase to low temperature metallic/semi-metallic phase in the present Ag$_2$Se sample appears to be linked to the temperature induced ETT. 

To show that the present Ag$_2$Se sample undergoes a temperature induced first-order electronic topological transition, we plot in Fig.4, the results of UPS on Ag$_{2}$Se. The UPS of Ag$_{2}$Se at 300~K and below are measured using 21.2~eV photons from HeI Source and the experimental end station (SPECS GmbH, Germany) at Indus-2  \cite{som20}. In this case, the energy resolution is limited mainly by temperature. The UPS at 300~K and above are measured using AIPES beamline at Indus-1 Synchrotron Source \cite{cha02} with a energy resolution of about 0.2~eV. The UPS measured at (i) 300~K before cooling (300~K (BC)), (ii)  15~K and (iii) 300~K after warming from 15~K (300~K (AW)) are shown in Fig. 4 (a). The gross features of the spectra match with that reported in the literature \cite{qu18}. The valence band spectra have four features which are marked as A, B, C and D in Fig. 4(a). The large density of states at feature A in the binding energy range 4.5-7.5~eV corresponds to the Ag $d$ -states \cite{ali19, fan02}.  All the spectra have been normalized to the intensity at the peak position. A small hump B at 3.5~eV appears at 15~K, which was absent before cooling. This hump does not disappear completely when the sample is heated again to 300~K. This irreversibility indicates that the modification in the density of states is responsible for the observed transport properties. The band edge states (C) at 1~eV are derived mainly from the Se $p$ states \cite{ali19, fan02}. At the Fermi level ($E_F$ = 0~eV), the UPS at 300~K (BC) shows the absence of electronic states in the range from $E_F$ to 0.4~eV below $E_F$ confirming the semiconducting behaviour. The spectra is similar to that reported in the literature\cite{qu18}. It may be noted that the energy gap estimated from the $\rho$($T$) and $n$($T$) is quite low. This is because of the presence of in-gap states originating from from the disorder which is quite strong in the present sample. These states contribute to the background in the UPS spectra. When the sample is cooled to 15~K, a clear Fermi edge is seen. This indicates that the sample has transformed to a metallic/semi-metallic state when cooled below 300~K. The Ag 4$d$ peak at 5.18~eV observed at 300~K (BC) shifts to 5.32~eV at 15~K. This indicates that the Fermi level moves 0.14~eV towards the conduction band. In other words, a conduction band crossed the Fermi level when temperature is reduced below 300~K. The $n$ shown in Fig. 1(b) decreases exponentially with decreasing temperatures below 395~K indicting $n$-type conduction. Below $\approx$250~K, the $n$ is nearly constant revealing that the sample is in the metallic state. There is no change in the sign of the Hall coefficient which again indicates that the metallic state appears due to the fact that the Fermi level lies in the conduction band. Figure 4(b) shows the evolution of band structure around the Fermi energy when the temperature is varied between 395~K to 15~K. At 395~K, the Ag$_2$Se shows a gapped state.  When the temperature is reduced below 300~K, the electronic states start to appear at the Fermi level. This is indicated by a finite intensity at $E_F$ binding energy in the temperature range 260-15~K. During heating, the density of states at $E_F$ persists till 300~K (for comparison, the data obtained from both the experimental setups are plotted for 300~K). The Fermi edge completely disappeared after heating to 330~K and above. Thus, we confirm that Ag$_2$Se undergoes a temperature induced first-order electronic topological transition.           

In conclusion, we have shown that there is a hysteresis in the temperature dependence of the physical properties of Ag$_2$Se. The hysteresis is observed to be of electronic origin. We conclude that the observed properties of Ag$_2$Se is due to a temperature induced first order electronic topological transition. This transition from high-$T$ semiconducting to low-$T$ metallic state over a large temperature range leads to a constant and moderately high value of $ZT$ in the temperature range 300-395~K. It is also to be noted that the $ZT$ variation during heating and cooling cycle does not change much.  On the other hand, using Ag$_2$Se for high magnetic field sensor application requires careful considerations as the absolute value of resistivity at any temperature above 100~K during heating cycle is significantly different from that while cooling. 

{\it Acknowledgement:} We thank Mr. Sharad Verma, UGC-DAE CSR, Indore, for his help in the PES experiments. We also thank Dr. A. K. Sinha, RRCAT, Indore and Dr. T. Ganguli, RRCAT, Indore as well as Dr. D. M. Phase, UGC-DAE CSR, Indore, for providing XRD and PES facilities respectively. 

{\it Data Availability Statement:} The data that support the findings of this study are available from the corresponding author upon reasonable request.

\end{document}